\documentclass{IEEEtran}
\usepackage{array}
\usepackage{graphicx, subfigure, color}
\usepackage{float}
\usepackage{subfloat}
\usepackage{multirow}
\usepackage{rotating}
\usepackage{amsmath} 
\usepackage{amssymb}  

\begin{document}

\title{Frames in Outdoor 802.11 WLANs Provide a Hybrid Binary-Symmetric/Packet-Erasure Channel }
\author{Xiaomin Chen, Douglas Leith, \\Hamilton Institute, NUI Maynooth\thanks{Supported by Science Foundation Ireland grants 07/IN.1/I901, 11/PI/11771. }}
\maketitle

\begin{abstract}
Corrupted frames with CRC errors potentially provide a useful
channel through which information can be transmitted.  Using
measurements taken in an outdoor environment, it is demonstrated
that for 802.11 wireless links the channel provided by corrupted
frames alone (i.e. ignoring frames with PHY errors and frames
received correctly) can be accurately modelled as a binary
symmetric channel (BSC) provided appropriate pre- and post-
processing is carried out.  Also, the channel provided by
corrupted frames and other frames combined can be accurately
modelled as a hybrid binary-symmetric/packet-erasure channel.
Importantly, it is found that this hybrid channel offers capacity
increases of more than 100$\%$ compared to a conventional packet
erasure channel over a wide range of RSSIs. This indicates that
the potential exists for significant network throughput gains if
the information contained in 802.11 corrupted packets is
exploited.
\end{abstract}

\section{Introduction}

\begin{figure}
 \centering
  \includegraphics[width=0.9\columnwidth,height=4cm]{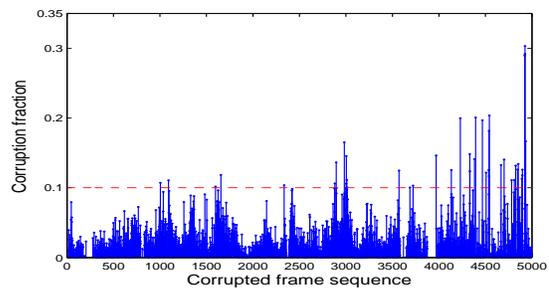}
  \caption{Measured fraction of incorrect bits vs frame sequence number.   Outdoor measurements, 802.11g PHY rate 54Mbps. $FER=91.98\%$. }\label{Crptfrac54M9198}
\end{figure}

Frames sent over an 802.11 wireless link may be received (i) with
a \emph{PHY error}, where the PHY header is corrupted by
noise/interference and the receiver cannot demodulate the frame,
or (ii) with a \emph{CRC error}, where the PHY header is received
correctly and the frame is decoded but then fails a CRC check or
(iii) \emph{without error}.   The resulting information channel is
often modelled as a packet erasure channel.  That is, frames are
received in an ``all or nothing'' fashion with frames having PHY
or CRC errors being discarded and only frames received without
error retained. However, the fraction of incorrect bits in frames
with CRC errors can be small. For example,
Fig.~\ref{Crptfrac54M9198} shows measurements of the
fraction of corrupted bits in frames with CRC errors on an 802.11
link.  It can be seen that even when the frame error rate (FER)
is high ($91.98\%$ of frames fail the PHY header check or the CRC
check), most of the frames received with CRC errors have less than
$10\%$ of bits incorrect.  Thus, these corrupted frames
potentially provide a useful channel through which we can transmit
information.

Taking this observation as our starting point, in this paper our
aim is to characterise the information channel provided by 802.11
frame transmissions.  Using measurements taken in an outdoor
environment,  we demonstrate that the channel provided by
corrupted frames alone (i.e. ignoring frames with PHY errors and
frames received correctly) can be accurately modelled as a BSC
provided appropriate pre- and post- processing is carried out.
Also, the channel provided by corrupted frames and other frames
combined can be accurately modelled as a hybrid
binary-symmetric/packet-erasure channel. We calculate the capacity
of this channel and show that the potential exists for significant
throughput gains.   {This complements and extends recent work
\cite{ppr,ziptx,maranello} which study the gains achievable in
802.11 by use of a refined erasure model (partitioning frames into
blocks and adding checksums).}

\section{Preliminaries}
\subsection{Experimental setup}\label{sec:setup}
Experimental data was collected in an open outdoor space with no
other interferers present. The FER was adjusted by varying the
distance between sender and receiver. Care was taken to ensure
repeatability of results -- measurements were taken on an open
space (a large playing field), sender and receiver were positioned
at fixed heights, antenna orientations were held fixed, human
operators left the vicinity of the experiment during measurements.

\subsubsection{Hardware and software}\label{measurementsetup}
An Asus Eee PC 4G Surf equipped with Atheros AR5BXB63 802.11b/g
chipsets (AR2425, MAC 14.2, RF5424, PHY 7.0)  was used as the
access point, running FreeBSD 8.0 with the RELEASE kernel and
using the standard FreeBSD ATH driver. A Fujitsu E series Lifebook
equipped with a Netgear dual band 802.11a/b/g  wireless PC card
WAG511 using Atheros AR5212 chipset was used as a client station,
running Ubuntu 11.04  and using a modified Linux Madwifi driver.
Onmi antennas used.

We disabled the Atheros' Ambient Noise Immunity feature which has
been reported to cause unwanted side effects.
Transmission power of the laptops was fixed and antenna diversity
disabled.   In previous work we have taken considerable care to
confirm that with this hardware/software setup the wireless
stations accurately follow the IEEE 802.11 standard, see
\cite{txop_technique}.

A perl script was used to generate CBR UDP traffic. The content of
each UDP packet was a random binary vector.  Unless stated
otherwise, the UDP payload is $8000$ bits, and the inter-packet
interval is $20$ ms. Packets were transmitted over the WLAN in the
broadcast mode and hence there were no MAC level ACKs or
retransmissions. The wireless driver at the receiver was modified
to record the receiving status for frames with PHY errors, CRC
errors or without errors and to transfer the contents of these
from the kernel to user space via the high-speed data relay
filesystem (relayfs).

\subsubsection{Recovering sequence number for corrupted frames}

Frames received with CRC errors were compared against the
corresponding original content in order to determine which
specific bits inside the frame were received corrupted. Since the
frame header might also be corrupted, the following fitting and
pattern matching procedure was used to indirectly recover the frame
sequence number.  For each corrupted frame we first searched the set
of correctly received frames to find the packet received without error
closest in time to the corrupted frame. Since the frames are
transmitted at a fixed rate, the interval between two consecutive
frame timestamps is roughly constant, but to correct for clock
skew between transmitter and receiver it is necessary to estimate
the relative clock rate and offset.   We estimated these using
linear least squares fitting of the received timestamps of
neighbouring error-free frames.    The  timestamp of the
corrupted frame was then used to determine likely candidates
for this frame amongst the transmitted frames.   From this set of
candidates, the sequence number and payload of each was compared
with that of the corrupted frame in order to identify the transmitted frame
most likely to correspond to the received corrupted frame.

\subsection{Runs test}

Our statistical analysis makes use of the \emph{runs test} (also
called the Wald-Wolfowitz test)~\cite{runstest}. The runs test is
a non-parametric test to check the null hypothesis that the
elements in a two-valued sequence are independent and identical
distributed. Given a 0-1 sequence, a \emph{run} is consecutive
entries having the same value \emph{e.g.} in the sequence
$1100110111$ there is one 0 run, one 00 run, two 11 runs and one
111 run. Under the null hypothesis, the number of runs is a random
variable whose conditional distribution is approximately normal
with mean $\mu=\frac{2\ N_1\ N_0}{N}+1$ and variance
$\sigma^2=\frac{(\mu-1)(\mu-2)}{N-1}$ where $N_1$ is the number of
1 values in the sequence, $N_0$ the number of 0 values and $N =
N_1 + N_0$.   Unless otherwise stated we carry out statistical
testing at the 5$\%$ significance level.

\section{Channel modelling}

We proceed by first investigating the channel provided by
corrupted frames alone \emph{i.e.} ignoring frames with PHY errors
and frames received correctly.   We find that, to within
statistical error, this can be accurately modelled as a BSC. We
then consider the channel provided by corrupted frames and other
frames combined.  We find that this can be accurately modelled as
a hybrid binary-symmetric/packet-erasure channel.

\subsection{Channel provided by corrupted frames}\label{sec:bsc}

A binary symmetric channel (BSC) takes binary input $X\in\{0,1\}$ and
maps this to binary output $Y\in\{0,1\}$.   With probability $1-p$
the channel  transmits the input bit correctly and with crossover
probability $p$ the input bit is flipped.  That is,
$\mathbb{P}\{Y=1|X=1\}=1-p=\mathbb{P}\{Y=0|X=0\}$ and
$\mathbb{P}\{Y=0|X=1\}=p=\mathbb{P}\{Y=1|X=0\}$. Repeated binary channel
uses are independent and identically distributed. That is, if a
binary vector is transmitted through a BSC, each bit is flipped
independently and with identical crossover probability $p$.
Therefore, to establish that a channel taking binary inputs is a
BSC, we need to show\newline
\noindent(1) Repeated binary channel uses are independent and identically distributed (i.i.d.).\newline
\noindent (2) The probability that a $1$ is flipped to a $0$ after
transmission is the same as the probability that a $0$ is flipped
to a $1$ i.e. the binary channel is symmetric.

\subsubsection{Binary channel uses are i.i.d.} \label{iidtest}
\begin{figure}
  \centering
  \subfigure[Before interleaving]{\label{BERbefore}\includegraphics[height=2.5cm, width=0.9\columnwidth]{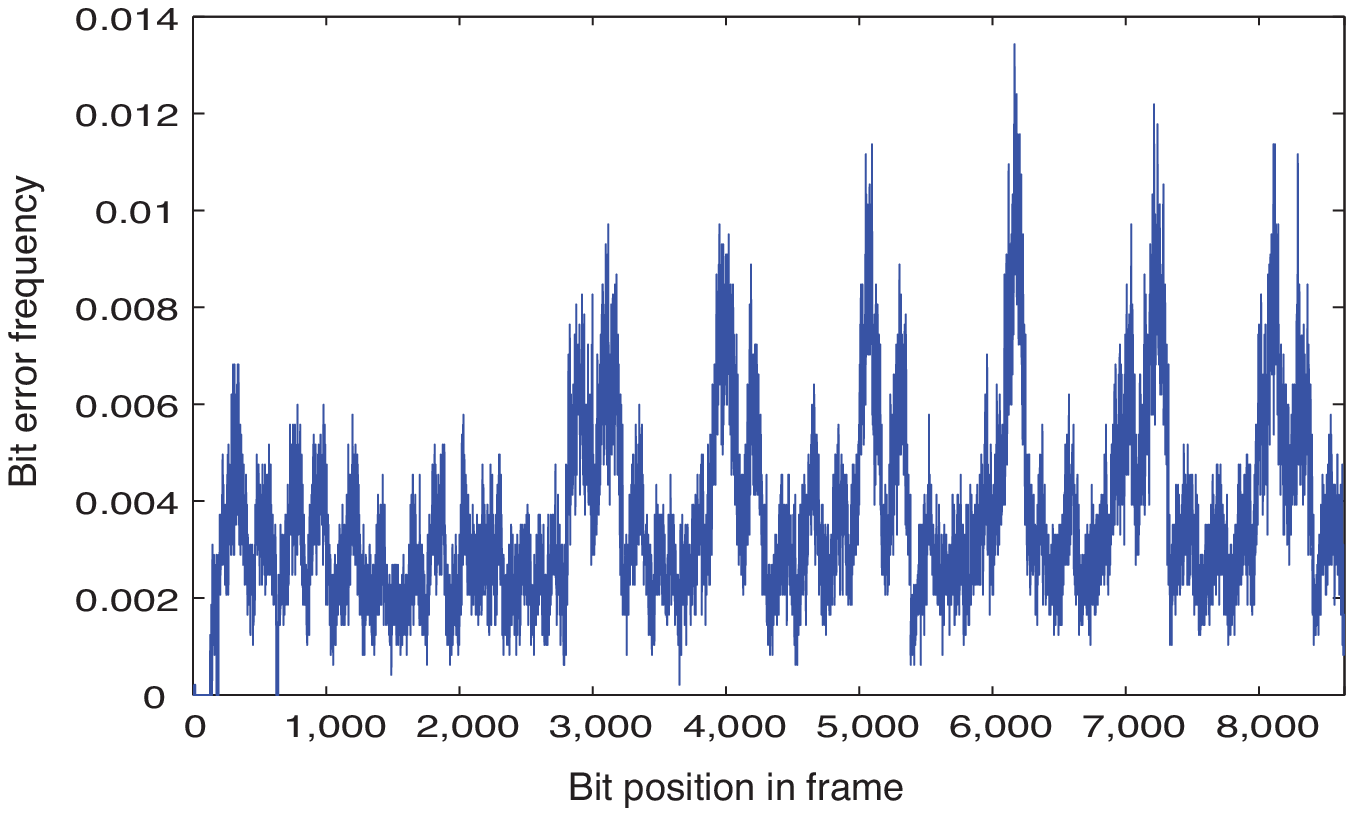}}
  \subfigure[After interleaving]{\label{BERafter}\includegraphics[height=2.5cm, width=0.9\columnwidth]{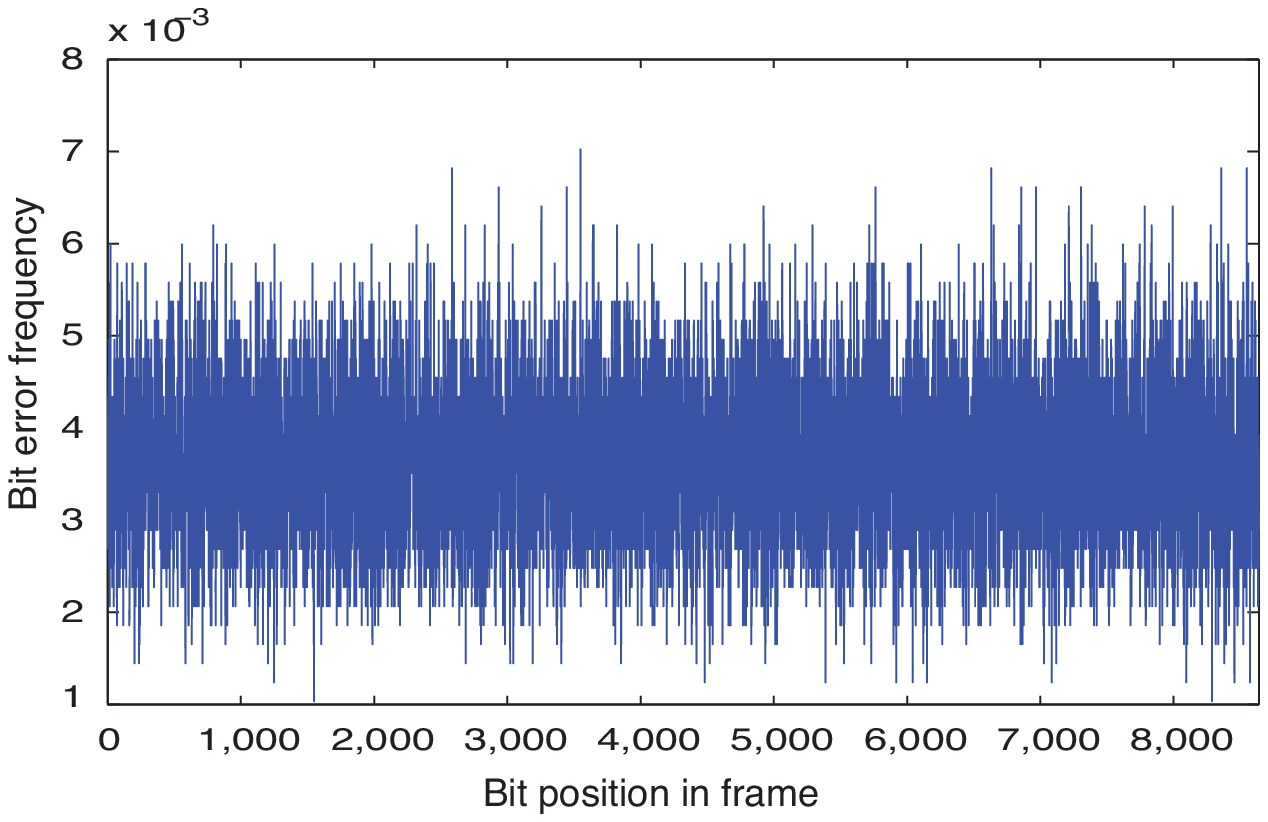}}
  \caption{Per-bit error frequency pattern across a frame before and after interleaving, outdoor, $FER=0.5658$, PHY rate 54Mbps, 5000 frames.}
  \label{fig:animals}
\end{figure}

We begin by presenting raw experimental measurements in Fig.
\ref{BERbefore}.  This figure plots representative measurements of
the bit error frequency for each bit position within a corrupted
frame.  It can be seen that bit errors are not evenly distributed
and the bit error frequency sequence is periodic across a frame.
This observation is not new, \emph{e.g.} see \cite{infocom09,
Paul, industrial}, and clearly violates the independence
requirement of a BSC. Nevertheless, Fig.~\ref{BERafter} plots the
bit error frequency for the same data after {interleaving}
\footnote{ That is, we randomly permute the bits within each
frame.  While the 802.11 PHY already carries out interleaving,
this is carried out over small blocks of bits equal in size to an
OFDM symbol, e.g. 288 bits with 64-QAM, whereas we interleave over
a complete frame i.e 8000 bits when the frame size is 1000B. Note
that interleaving does not maximise capacity.}. Interleaving can
be readily implemented at the MAC layer -- bits in a frame are
permuted at the transmitter before a MAC frame goes down to the
PHY layer and when the frame is received the inverse permutation
is used to recover the original bit order.   It can be seen from
Fig. \ref{BERafter} that  after interleaving the periodicity of
bit errors appears to be removed, although further analysis is
required to confirm this.

To analyse the independence of repeated channel uses within each
individual interleaved frame we use the runs test.   For each
corrupted frame we construct a 0-1 sequence by labelling corrupted
bits as $1$'s and correct bits as $0$'s.   We find that the runs
test cannot reject the null hypothesis at the 5$\%$ significance
level that after interleaving bit errors inside a frame are
independent. That is, to within statistical error we can conclude
that repeated channel uses inside each interleaved frame are
independent and Bernoulli distributed with the bit crossover
probability $p=n_{c}/l$, where $n_{c}$ is the number of corrupted
bits within a frame and $l$ is the frame length.

\begin{figure}
 \centering
  \includegraphics[height=3.5cm, width=0.9\columnwidth]{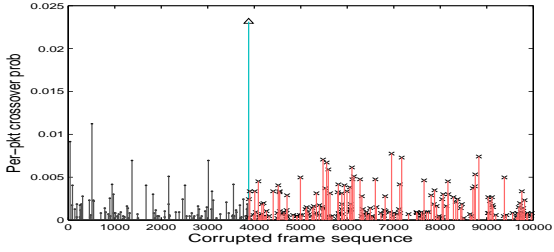}
  \caption{Per-frame bit crossover probability for a sequence of 10000 packets, PHY rate 54Mbps, $FER=0.0423$. }\label{54M0423}
\end{figure}

\begin{figure}
 \centering
  \includegraphics[height=4cm, width=0.9\columnwidth]{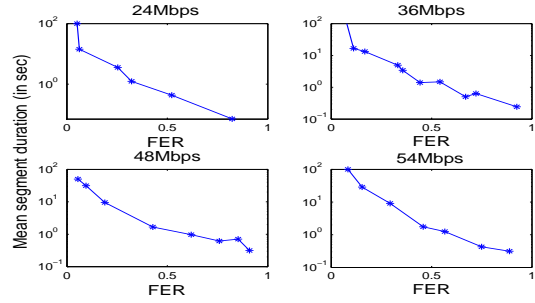}
  \caption{Mean segment duration versus FER at different PHY rates. }\label{SegDuration}
\end{figure}

Next, to analyse the independence of channel uses across multiple
frames we concatenate the foregoing binary sequences for
successive corrupted frames and apply the runs test. For example,
Fig.~\ref{54M0423} plots the measured per-frame bit crossover
probability for a sequence of 10,000 frames.  Consecutive
corrupted frames that pass the runs test are labelled using the
same marker. It can be seen that the experimental run is
partitioned into three segments. Within each segment, corrupted
frames form a bit-level channel over which repeated channel uses
can be considered to be independent and Bernoulli distributed to
within statistical error. The first segment spans 3865 frames
(around 77.3s given that the inter-packet interval is 20ms) and
the third segment spans 6118 frames (about 122.36s). The second
segment has only one frame with an unusually high crossover
probability.    Neglecting this single bad frame out of 10,000
frames, the measurement results indicate that for long periods the
sequence of corrupted frames form a bit-level channel over which
repeated channel uses can be considered to be i.i.d. The data in
Fig.~\ref{54M0423} is for a FER of 0.0423 and PHY rate of 54Mbps.
Fig.~\ref{SegDuration} plots the mean of segment durations over an
experimental run of 10000 packets for a range of FERs and PHY
rates. It can be seen that the segment duration tends to decrease
as the FER increases \emph{i.e.} the duration within which
corrupted frames have an i.i.d. bit crossover probability becomes
shorter as the FER increases. For FER's less than 20\%, at all PHY
rates the mean segment duration exceeds 10s. That is, for FER's
less than 20\% the binary channel provided by corrupted frames is
i.i.d. for periods exceeding, on average, 10s. Such time-scales
seem sufficient for most channel modelling purposes.

\subsubsection{Binary channel is symmetric}

Table~\ref{errorrate} reports the measured bit flip rates for $1$
values and $0$ values.  Results are shown for a range of FER
values and PHY rates. It can be seen that the bit flip
probabilities for both $1$'s and $0$'s are close to each other ,
and thus, to within experimental error, can be approximately
considered as ``symmetric''.
\begin{table*}
\scriptsize
\centering
\begin{tabular}{|c|c|c|c|c|c|c|c|c|c|c|c|}
\hline
\multirow{2}{*}{PHY rate} &  \multirow{2}{*}{FER}  & \multicolumn{2}{|c|}{Flip rate for $1$'s}& \multicolumn{2}{|c|}{Flip rate for $0$'s }& \multirow{2}{*}{PHY rate} &  \multirow{2}{*}{FER}  & \multicolumn{2}{|c|}{Flip rate for $1$'s }& \multicolumn{2}{|c|}{Flip rate for $0$'s }\\
\cline{3-6}\cline{9-12}
& & $\mu_1$ & $\sigma_1/\sqrt{N_1}$ & $\mu_0$ & $\sigma_0/\sqrt{N_0}$ & & & $\mu_1$ & $\sigma_1/\sqrt{N_1}$ & $\mu_0$ & $\sigma_0/\sqrt{N_0}$ \\
\hline \multirow{8}{*}{54Mbps}
 &0.0835 & 0.0018 & 2.34$\times10^{-5}$ & 0.0018 & 2.26$\times10^{-5}$ &
 \multirow{8}{*}{36Mbps} &0.0649 & 0.0014 & 2.72$\times10^{-5}$ & 0.0013 & 2.48$\times10^{-5}$ \\
 &0.0984 & 0.0019 & 2.40$\times10^{-5}$ & 0.0020 & 2.34$\times10^{-5}$ & & 0.1673 & 0.0019 & 1.69$\times10^{-5}$ & 0.0020 & 1.65$\times10^{-5}$\\
 &0.1540 & 0.0023 & 1.97$\times10^{-5}$ & 0.0023 & 1.88$\times10^{-5}$ & & 0.3314 & 0.0023 & 1.31$\times10^{-5}$ & 0.0023 & 1.22$\times10^{-5}$\\
 &0.2932 & 0.0031 & 1.67$\times10^{-5}$ & 0.0032 & 1.62$\times10^{-5}$ & & 0.4427 & 0.0032 & 1.34$\times10^{-5}$ & 0.0033 & 1.30$\times10^{-5}$\\
 &0.4584 & 0.0042 & 2.19$\times10^{-5}$ & 0.0041 & 2.04$\times10^{-5}$ & & 0.5416 & 0.0037 & 1.19$\times10^{-5}$ & 0.0035 & 1.10$\times10^{-5}$\\
 &0.5658 & 0.0067 & 2.45$\times10^{-5}$ & 0.0065 & 2.31$\times10^{-5}$ & & 0.6697 & 0.0052 & 1.43$\times10^{-5}$ & 0.0054 & 1.38$\times10^{-5}$\\
 &0.7492 & 0.0066 & 1.52$\times10^{-5}$ & 0.0068 & 1.48$\times10^{-5}$ & & 0.7214 & 0.0052 & 2.27$\times10^{-5}$ & 0.0048 & 2.08$\times10^{-5}$\\
 &0.8881 & 0.0120 & 1.92$\times10^{-5}$ & 0.0121 & 1.83$\times10^{-5}$ & & 0.9247 & 0.0122 & 3.41$\times10^{-5}$ & 0.0115 & 3.15$\times10^{-5}$\\
\hline \multirow{8}{*}{48Mbps}
& 0.0547 & 0.0015 & 3.21$\times10^{-5}$ & 0.0017 & 3.16$\times10^{-5}$ & \multirow{4}{*}{24Mbps} &  0.0617 & 0.0048 & 7.66$\times10^{-5}$ & 0.0044 & 6.99$\times10^{-5}$\\
 &0.0949 & 0.0021 & 2.91$\times10^{-5}$ & 0.0020 & 2.71$\times10^{-5}$ & & 0.3219 & 0.0041 & 1.84$\times10^{-5}$ & 0.0039 & 1.73$\times10^{-5}$\\
 &0.1883 & 0.0021 & 1.68$\times10^{-5}$ & 0.0020 & 1.59$\times10^{-5}$ & & 0.5220 & 0.0098 & 2.68$\times10^{-5}$ & 0.0094 & 2.50$\times10^{-5}$\\
 &0.4319 & 0.0034 & 1.43$\times10^{-5}$ & 0.0034 & 1.35$\times10^{-5}$ & & 0.8229 & 0.0306 & 3.19$\times10^{-5}$ & 0.0293 & 2.97$\times10^{-5}$\\
 \cline{7-12}
 &0.5818 & 0.0036 & 1.81$\times10^{-5}$ & 0.0035 & 1.70$\times10^{-5}$ & \multirow{5}{*}{18Mbps} & 0.0449 & 0.0045 & 1.44$\times10^{-5}$ & 0.0046 & 1.38$\times10^{-5}$\\
 &0.6185 & 0.0044 & 1.38$\times10^{-5}$ & 0.0046 & 1.32$\times10^{-5}$ & & 0.3481 & 0.0049 & 2.15$\times10^{-5}$ & 0.0050 & 2.07$\times10^{-5}$\\
 &0.7586 & 0.0045 & 1.76$\times10^{-5}$ & 0.0043 & 1.64$\times10^{-5}$ & & 0.6415 & 0.0076 & 1.99$\times10^{-5}$ & 0.0075 & 1.89$\times10^{-5}$\\
 &0.8522 & 0.0080 & 2.30$\times10^{-5}$ & 0.0077 & 2.15$\times10^{-5}$ & & 0.8376 & 0.0091 & 1.93$\times10^{-5}$ & 0.0091 & 1.84$\times10^{-5}$\\
 &0.9091 & 0.0146 & 2.09$\times10^{-5}$ & 0.0145 & 1.98$\times10^{-5}$ & & 0.9776 & 0.0129 & 3.79$\times10^{-5}$ & 0.0123 & 3.53$\times10^{-5}$\\
\hline
\end{tabular}
\caption{Bit flip rates for $1$'s and $0$'s, $\mu_i$ the mean flip
rate of bit $i$ , $\sigma_i/\sqrt{N_i}$ the standard deviation of
flip rate of bit $i$ , $N_i$  the total number of bit $i$ in
corrupted frames, $N=N_0+N_1$.}\label{errorrate}
\end{table*}

In summary, after interleaving the bit errors in corrupted frames
are, to within statistical error, independent and identically
distributed with a symmetric crossover probability, \emph{i.e.}
the information channel can be
accurately modelled as a BSC.

\subsection{Hybrid binary-symmetric/packet-erasure channel}

\begin{figure}
 \centering
  \includegraphics[height=2cm,width=0.3\columnwidth]{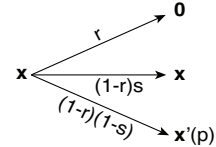}
  \caption{Hybrid BSC/packet erasure channel model.}\label{NewModel}
\end{figure}

Our hypothesis is that the channel provided by 802.11 frames can
be accurately modelled as a mixed packet erasure/binary symmetric
channel.  Formally, the channel takes an $n$-bit  binary vector
$\boldsymbol{x}\in\{0,1\}^{n}$ as input, and outputs received
vector $\boldsymbol{y}\in\{0,1\}^{n}$. The vector $\boldsymbol{y}$
can be received with three possible states: (i) $\boldsymbol0$
(erased), (ii) $\boldsymbol x'$ (corrupted) and (iii) $\boldsymbol
x$ (without error).  Repeated channel uses are i.i.d.  The
probability that a received vector is erased is $
\mathbb{P}\{\boldsymbol{y}=\boldsymbol{0}\}=r$. The probability
that a non-erased vector is received without error is
$\mathbb{P}\{\boldsymbol{y}=\boldsymbol{x}\ |\ \boldsymbol{y}\neq
\boldsymbol0\}=s$ and so the probability that a vector is
correctly received is
$\mathbb{P}\{\boldsymbol{y}=\boldsymbol{x}\}= (1-r)s$. The
probability that a vector is corrupted is
$\mathbb{P}\{\boldsymbol{y}=\boldsymbol{x}'\}=(1-r)(1-s)$. In a
corrupted vector bits are flipped symmetrically and independently
with crossover probability $p$. This channel is illustrated
schematically in Fig.~\ref{NewModel}.

We can relate this hybrid model to 802.11 by associating the input
vectors with transmitted frames, erasures with PHY errors and
corrupted vectors with CRC errors.  To establish an equivalence we
need to show that (i) frames are received with PHY errors, CRC
errors and without error in an i.i.d. fashion, (ii) corrupted
frames provide a BSC.  We have already established (ii) in Section
\ref {sec:bsc} but it remains to establish (i).

\begin{figure}
 \centering
  \includegraphics[height=4.8cm,width=\columnwidth]{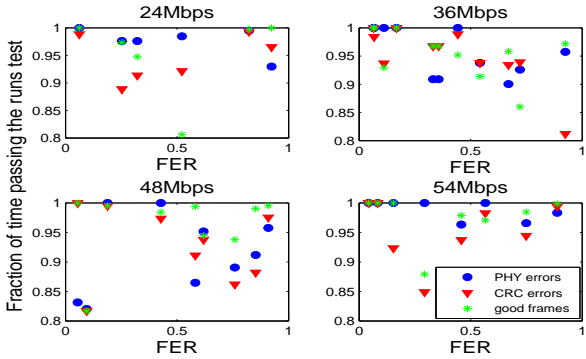}
  \caption{Fraction of time passing the runs test for PHY errors, CRC errors and good frames.}\label{passrate}
\end{figure}

Our hypothesis is that PHY errors, CRC errors and frames received
without errors are mutually independent across time. To
investigate this hypothesis, we again use the runs test. We
construct a 0-1 sequence by labelling frames received with PHY
errors as a $1$ and other frames as a $0$, and then apply the runs
test for each individual segment which passes the runs test for
bit error independence check. Similarly, we do the same
respectively for frames received with CRC errors and without
errors. Fig.~\ref{passrate} plots the fraction of time in an
experimental run of 10,000 frames within which the runs tests pass
for a range of FERs and PHY rates. The runs
tests pass over at least 80$\%$ of the time, \emph{i.e.} in a run of 10,000 frames over
80$\%$ of the time, to within statistical error, frames are
received with PHY errors, CRC errors and without error in an
i.i.d. fashion.

\section{Conclusions}

\begin{figure}
 \centering
  \includegraphics[height=4.8cm, width=0.9\columnwidth]{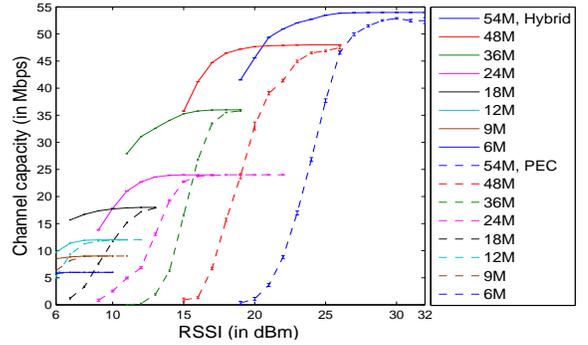}
  \caption{Outdoor experimental channel capacity. }\label{outdoorchannelcapacity}
\end{figure}

The capacity of the hybrid channel is
\begin{equation}
C=R(1-r)\left(s+(1-s)\big(1-H(p)\big)\right)
\end{equation}
where $r$ is the probability of a PHY error, $s$ is the
probability that a non-erased packet is error free, $p$ is the
crossover probability in corrupted packets,
$H(p)=-p\log_2(p)-(1-p)\log_2(1-p)$ and $R$ is the PHY bit rate.
Using our experimental data, we binned frames according to their
RSSI (Received Signal Strength Indication) and calculated the
measured crossover probability in corrupted frames and the frame
error rate.  As we could not measure the RSSI for PHY erasures
using our hardware,  we do not include these here.
Fig.~\ref{outdoorchannelcapacity} shows the resulting measured
capacity vs RSSI.   For comparison we also plot the experimentally
measured packet erasure channel capacity \emph{i.e.} the capacity
when corrupted frames are discarded.  It is important to note that
these measured curves cannot be directly compared with theoretical
calculations since the mapping between RSSI and SNR is not well
defined. It can be seen that the capacity of the hybrid channel is
strictly greater than the erasure channel capacity, as expected.
More interestingly, it can be seen that the hybrid channel offers
capacity increases of more than 100$\%$ over a wide range of
RSSIs. This indicates that the potential exists for significant
network throughput gains if the information contained in corrupted
packets is exploited. Since 802.11n includes the 802.11a/g
modulation and coding schemes, our conclusions apply directly.  We
do, however, expect that indoor links will behave differently from
outdoor links due to multipath effects and temporal variations in
the environment, and this will be the subject of future work.

{}

\end{document}